\documentclass[%
 aip,
jcp,
 jmp,%
 amsmath,amssymb,
preprint,%
]{revtex4-1}

\usepackage{graphicx}
\usepackage{dcolumn}
\usepackage{bm}

\newcommand{\ket}[1]{|{#1}\rangle} 
\begin{document}


\title{Sensitivity test of a blue-detuned dipole trap designed for parity non-conservation measurements in Fr}

\author{D. Sheng}
\altaffiliation[Present address: ]{Department of Physics, Princeton University, Princeton, NJ 08544, U. S. A.}
\affiliation{Joint Quantum Institute, Department of Physics, University of Maryland, and National Institute of Standards and Technology, College Park, MD 20742, U. S. A.}

\author{J. Zhang}%
\affiliation{Joint Quantum Institute, Department of Physics, University of Maryland, and National Institute of Standards and Technology, College Park, MD 20742, U. S. A.}

\author{L. A. Orozco}
 \email{lorozco@umd.edu}
 \homepage{http://www.physics.umd.edu/rgroups/amo/orozco/index.html}
\affiliation{Joint Quantum Institute, Department of Physics, University of Maryland, and National Institute of Standards and Technology, College Park, MD 20742, U. S. A.}

\date{\today}

\begin{abstract}
A dynamic blue-detuned optical dipole trap with stable $^{87}$Rb atoms produces a differential ac Stark shift of 18 Hz in the ground state hyperfine transition, and it preserves the ground state hyperfine superpositions for a long coherence time of 180 ms. The trapped atoms undergoing microwave Rabi oscillations are sensitive to a small signal, artificially generated with a second microwave source,
phase locked to the first allowing a simple and effective method for determining signal-to-noise ratio limits through interference techniques. This provides an excellent means of calibrating sensitivity in experiments such as our ongoing Fr parity non-conservation measurement.
\end{abstract}

\pacs{31.30.jg, 37.10.Gh, 29.90.+rj}
\keywords{Parity nonconservation, blue detuned dipole trap, weak interference}
\maketitle

\section{Introduction}
This paper presents a blue-detuned dipole trap \cite{grimm99} apparatus developed for precision measurements with an interference method to evaluate its sensitivity. A single focused laser beam dynamically rotated off axis forms the trap \cite{friedman00,rudy01},  and the atoms spend most of the time in the darkness, with minimal perturbations. Specifically,  we first quantify the differential ac Stark shift of the ground state hyperfine splitting of $^{87}$Rb atoms in the trap and measure the preservation of atomic coherence of the hyperfine superposition. Since the success of precision measurements relies on signal-to-noise ratio $(S/N)$, we show an interference method that uses a dual microwave horn arrangement to test {\it{in situ}} the response of the atoms to a perturbation comparable to that expected in precision measurements. Both the trap and the double horn interference method can have applications in ongoing and future tests of time (T)-violation and measurements of parity non-conservation (PNC) (see for example  \cite{chin01,romalis99,gomez07}).

Our group is interested in atomic PNC, particularly on the anapole moment observed in the nuclear-spin-dependent (NSD) part of atomic PNC. The anapole moment is the dominant contribution to NSD PNC in heavy atoms~\cite{zeldovich59,flambaum84,flambaum84b,ginges04}; it can be thought of as a weak radiative correction among nucleons probed by an electromagnetic interaction. 
There are two completed experiments on atomic PNC that have extracted anapole moment information. The one on Tl gives an error bound of the nuclear anapole moment~\cite{vetter95}, and the one on Cs finds a non-zero value with an error of about 15\%~\cite{wood97,wood99}, which has a similar uncertainty with other measurements in nuclear physics; however, when extracting meson coupling constants from these numbers, they do not agree with each other~\cite{haxton01, behr09}. Both atomic experiments measured the total PNC signal for different hyperfine states and compared the difference to extract the NSD PNC signal. New proposals and ongoing efforts to solve this discrepancy include ions \cite{fortson93,ferro11},  stable Yb atoms~\cite{tsigutkin09}, BaF molecules~\cite{demille08}, radioactive Fr atoms~\cite{gomez06,gomez07}, and a chain of Rb isotopes~\cite{sheng10}.

The anapole moment measurement~\cite{gomez07} benefits from a large nuclear charge ($Z$), as it grows roughly as  $Z^{8/3}$.
We are planning to use the radioactive element francium ($Z=89$), the heaviest alkali atom~\cite{aubin03a,gomez06}. Our experimental scheme  to measure the anapole moment  requires driving a resonant electric dipole $(E1)$ parity-forbidden transition between ground hyperfine states in a series of francium isotopes inside an optical dipole trap at the electric anti-node of a resonant microwave cavity. The measurement  interferes a parity conserving (PC) transition~\textendash~such as a magnetic dipole $(M1)$ transition~\textendash~with the PNC $E1$ transition under well-defined handedness to extract the small contribution from the weak interaction. Refs.~\cite{gomez07,sheng10} present a detailed study of the experimental requirements, including possible sources of systematic effects that can mimic the PNC signal. The francium experiment is currently on-going at the ISAC radioactive beam facility at TRIUMF in Vancouver, Canada. 
The first step in such a measurement is to develop a technique to assess and calibrate the sensitivity to the very small PNC signal that is expected. This paper presents such a technique and results with its use in stable $^{87}$Rb. 

\section{Sensitivity to PNC}
The states involved in the anapole moment experiment are all ground hyperfine states, and their lifetime does not limit the coherent interaction.  This is in contrast with the coherent interaction time in Ref.~\cite{wood97}, which is limited by the $7s$ excited state lifetime in Cs. The result is an improvement of the signal-to-noise ratio (S/N) per atom. Because the parity-violating transition amplitude ($A_{PV}$) is still too small to observe directly, we need to amplify the signal by interfering it with another coherent process between the same two states, a parity-conserved transition with a much larger transition amplitude ($A_{PC}$).

The electroweak interference requires the excitation of trapped francium atoms inside a microwave Fabry-Perot cavity. 
Three fields define the coordinate system of the apparatus: the microwave electric field from the cavity that drives $A_{PV}$, the static magnetic field aligned with the magnetic microwave field of the cavity, and an auxiliary microwave magnetic field aligned with the axis of the cavity that drives $A_{PC}$. In a geometry with the atoms confined to the anti-node of the electric microwave cavity field, only PNC $E1$ transitions between hyperfine levels are driven, while $M1$ transitions from the microwave cavity magnetic field are suppressed.

The electric dipole PNC transition amplitude $A_{PV}$ for $^{209}$Fr, between the hyperfine level
$\ket{1}=\ket{F=4,m_F=0}$ and $\ket{2}=\ket{F=5, m_F=-1}$ with a static magnetic field of 1553~Gauss and a microwave electric field {\bf E} of 476~V/cm, neglecting phase factors, is~\cite{gomez07} 
\begin{equation}
A_{PV}/\hbar=\Omega_{PV}= 0.01~{\rm{rad/s}}. \label{e1}
\end{equation}
The static magnetic field value minimizes the sensitivity to magnetic field fluctuations, and it is isotope-dependent.
The electric field amplitude requires cavity $Q$ factors on the order of $10^4$ with achievable microwave powers. We are currently working on the quasi-optical microwave cavity with a preliminary measurement of $Q=4\times10^4$.  

Next we estimate the sensitivity needed for the measurement. If we start with $N$ atoms in $\ket{1}$, the number of atoms $N_2$ ending in $\ket{2}$ after an interaction time $t_R$ is $N_2=N P_2$, where $P_2$ is the probability given by the Rabi oscillation of the population in $\ket{2}$:
\begin{eqnarray}
N_2&=&Ne^{-t_R/T_c} \times \nonumber \\
& & 
\sin^2\left(\frac{\left(A^2_{PC}+A^2_{PV}+2A_{PC}A_{PV}\cos\phi\right)^{1/2}t_R}{2\hbar}\right) +  \nonumber \\
& &\frac{N}{2}(1-e^{-t_R/T_C}),
\label{eq:intanapole}
\end{eqnarray}
where $\phi$ is the relative phase difference between these two transitions, and $T_C$ is the coherence time. By tuning the relative phase of these two transitions between $0$ and $\pi$ out of phase, we obtain a maximum change in the interference term. This change of $\pi$ in the relative phase relation is also equivalent to a coordinate reversal. The signal is the maximum change in the interference in the limit of small $A_{PV}$
\begin{eqnarray}
S&=&Ne^{-t_R/T_C}\sin\left(\frac{A_{PC}t_R}{\hbar}\right)\sin\left(\frac{A_{PV}t_R}{\hbar}\right) \nonumber \\  
&{\approx}&Ne^{-t_R/T_C}\sin\left(\frac{A_{PC}t_R}{\hbar}\right)\frac{A_{PV}t_R}{\hbar}.
\label{signal}
\end{eqnarray}
The signal in Eq.~\ref{signal} is linear in $A_{PV}$ of Eq.~\ref{e1}. For
a projection-noise-limited measurement~\cite{itano93} $ (N_P=\sqrt{N_2P_2(1-P_2)})$,  the signal-to-noise ratio is maximum when $P_2=0.5$, that is $\sin (A_{PC}t_R/\hbar)=1$. The signal-to-noise ratio after an interaction time   $t_R << t_C$ for $N$ atoms is
\begin{equation}
\frac{\ S}{{N}_P}=2\Omega_{PV}t_{R} \sqrt{N}.
\label{signaltonoise}
\end{equation}
This result requires a long coherence time and a large number of atoms to observe the weak interaction in a single shot. If we take  $\Omega_{PV}=0.01$ rad/s from Eq.~\ref{e1},  $t_R=0.05$ s, and $10^6$ atoms, we obtain a single shot $S/{N_P}=1$. Taking $n$ time averages increases the $S/N$ by $\sqrt{n}$, the same effect as the atom number $N$.

\section{Blue-detuned dipole trap and microwave system}
As discussed in the previous section, the atoms should be trapped with the minimum disturbance to their coherence properties. Our dipole trap aims to decrease the photon scattering and differential ac Stark shift introduced by the trapping laser. We use a far off-resonance trap (FORT) to reduce the photon scattering rate and choose a blue-detuned trap where the atoms are confined on the minima of the light field, the so-called dark region of the trap. The ac Stark shift depends on various parameters, including the position of atoms in the trap, the atomic state, and the time because the atoms move around inside the trap.

There are different optical configurations for generating blue-detuned traps. Our group has investigated the use of axicons, but diffraction creates avenues of escape~\cite{kulin01}. 
The efficiency of loading atoms from a cold cloud into the dipole trap depends on the volume of the trap region which calls for a large trap. On the other hand, the PNC experiment requires that the atoms be tightly trapped at the electric anti-node of the microwave cavity to reduce the unwanted $M1$ transition from the cavity magnetic microwave field. The dynamic trap provides a solution for initially trapping with a large volume and then compressing the trap dynamically~\cite{friedman00}. A laser rotating faster than the motion of the atoms creates a time-averaged potential equivalent to a hollow beam potential. The laser beam propagating in the $z$ direction goes through two acousto-optical modulators (AOM) (Crystal Technologies  3080-122) placed back-to-back in the $x$ and $y$ directions, respectively. The frequency tuning range of the AOMs is 20~MHz, with a center frequency of 80~MHz, which limits the trap size we generate.  We use the beam that corresponds to the first-order diffraction in both directions, the (1,1) mode. We scan the modulation frequency of both AOMs with two phase-locked function generators (Stanford Research Systems  DS345) to obtain different hollow beam shapes. Tightly focusing the laser at the position of the atoms confines them along the beam axis, which is perpendicular to the gravity direction. The cross section of the trap is a rhombus with a diagonal of 300 $\mu$m. The power in the dipole trap with 10.3 THz (20.8 nm) detuning is 530~mW, giving a maximum intensity of $2.1\times10^6$~mW/cm$^2$, and the polarization is perpendicular to the direction of gravity.  The well depth of the trap at this detuning is 24~$\mu$K, with atomic temperatures on the order of half the well depth. A detailed description of the apparatus is in Ref.~\cite{sheng10}.

\begin{figure}
\centering
\includegraphics[width=0.95\linewidth]{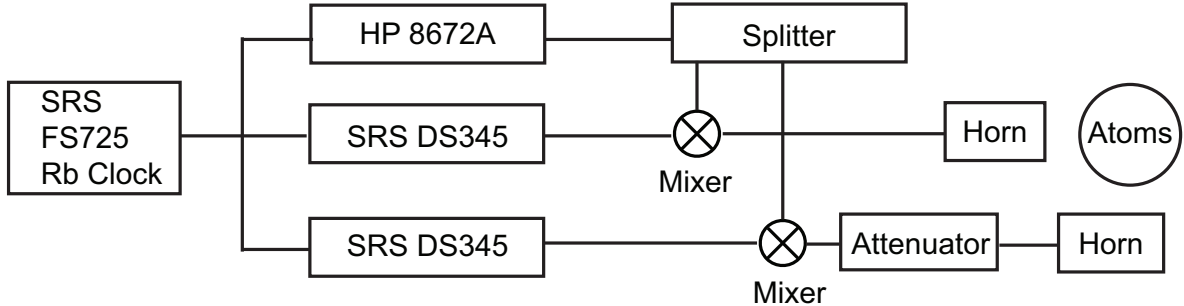}
\caption{\label{apparatus}  Schematic of the microwave system used for excitation of the trapped atoms. }
\end{figure}

Figure~\ref{apparatus} shows the schematic of the microwave part of the apparatus. A microwave source (HP 8672A) mixed with RF synthesizers (Stanford Research Systems DS345) and locked to a Rb atomic clock (Stanford Research Systems FR725) produces resonant excitation, with a few dBm of power, delivered through a microwave horn 24 cm away from the atoms. To test the interference scheme between a strong and very weak transition in our apparatus, we use the attenuated signal from a second horn.
We use the clock transition ($\ket{F=1,m_F=0}\rightarrow\ket{F'=2,m_F'=0}$) in the ground state of $^{87}$Rb to probe the coherence properties of the trap. We excite with microwaves to generate Rabi oscillations with a quantization magnetic field of 0.5~G in the direction of gravity. 

We drive the microwave transition for a set time, then we optically excite the atoms from a given hyperfine state and use a photomultiplier tube to detect the fluorescence which is proportional to the atom number in each state. We first measure the number of atoms in $\ket{5S,F=2}$ by driving a cycling transition to $\ket{5P_{3/2},F=3}$, then we turn on another laser beam on resonance with $\ket{5S,F=1}\rightarrow\ket{5P_{3/2},F=1}$  together with the previous cycling transition and measure the total number of atoms.

\section{Experimental results}

\subsection{Differential ac Stark shift}
The hyperfine splitting ($\hbar\omega_{HF}$) between the two ground states leads to a small difference in the ac Stark shift, which is the so-called differential ac Stark shift. When  $\delta$,  the laser detuning with respect to the atomic resonance, is large compared with $\omega_{HF}$, the differential ac Stark shift is 
$\Delta{U}(\textbf{r})=-U(\textbf{r})\omega_{HF}/\delta$, 
where $U(\textbf{r})$ is the dipole potential at the position of the atoms. Although the ac Stark shift has opposite signs for red- and blue-detuned dipole traps \cite{grimm99}, the differential shift is negative for both cases; it always decreases the hyperfine splitting. 
\begin{figure}
\centering
\includegraphics[width=0.95\linewidth]{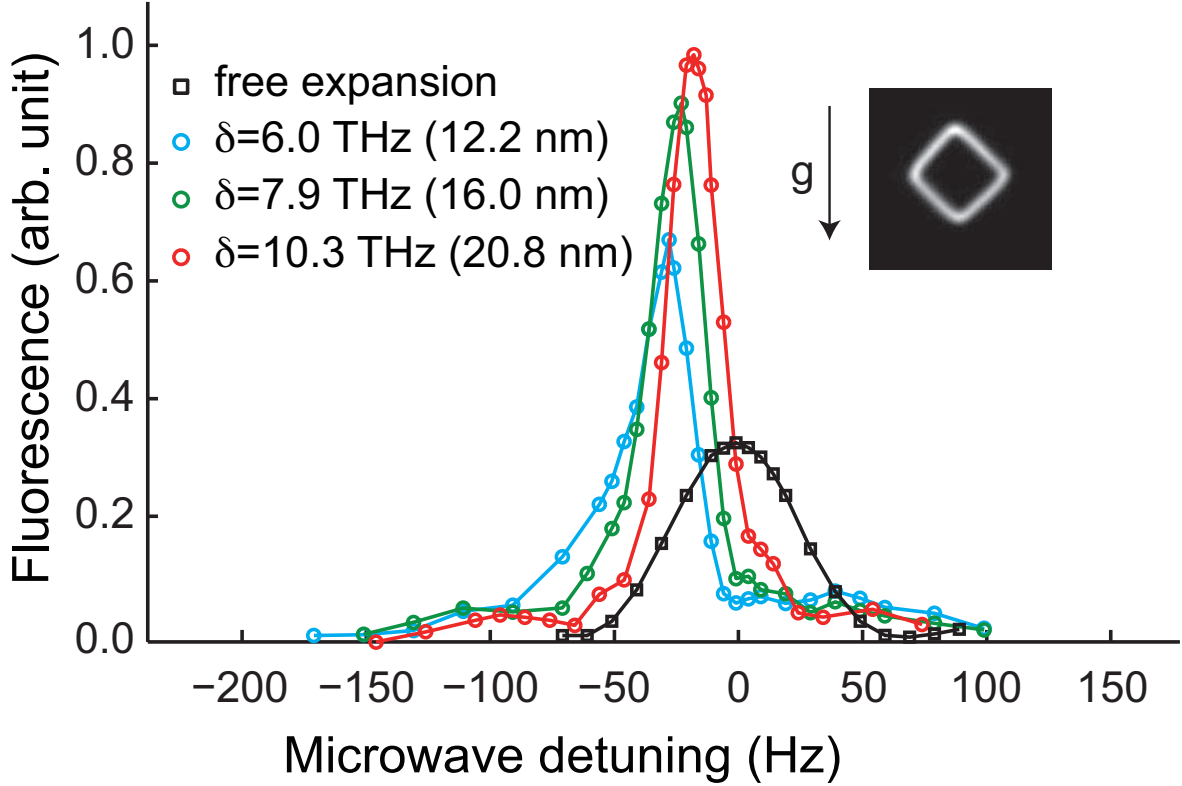}
\caption{\label{fig:lightshift}(Color online) Four examples of microwave resonances for different blue-detuned trap detunings. The differential ac Stark shift is clearly visible. Inset: shape of dipole trap with the arrow indicating the direction of gravity.}
\end{figure}

The differential ac Stark shift of atoms is simpler to map than the ac Stark shift~\cite{brantut08}. We first measure the unperturbed hyperfine splitting using cold atoms released from a magneto-optical trap (MOT) in the absence of the dipole trap (black symmetric trace in Fig.~\ref{fig:lightshift}). We have only 15~ms of interaction time because of gravity. When the dipole trap is on,  we have a longer interaction time $t_R $ that we choose as 40~ms, after which there are $N \approx 10^5$ atoms in the trap, and this interaction time limits the linewidth to about 20~Hz. 
The inset in Fig.~\ref{fig:lightshift} shows the cross section of the trap. The figure shows the differential shift for three detunings  while keeping the trapping beam power constant. The details of the linewidth and peak position of the differential ac Stark shift depend on the trap shape, light intensity, atomic energy, laser detunings, and atomic dynamics inside~\cite{sheng11}, which is quite different from the case of a red-detuned one~\cite{kim07}.

\subsection{Coherence of the ground state superposition}

The distribution of the differential ac Stark shift is a major source of inhomogenous broadening, and its linewidth determines the coherence time of atomic superposition in the trap~\cite{davidson95}. The data for the detuning at 10.3 THz ( 20.8~nm) from the $D_2$ line 
shows a differential shift of 18~Hz and a half linewidth of 10~Hz, very close to the observation-time broadening limit, which means a long coherence time.

\begin{figure}
\centering
\includegraphics[width=0.95\linewidth]{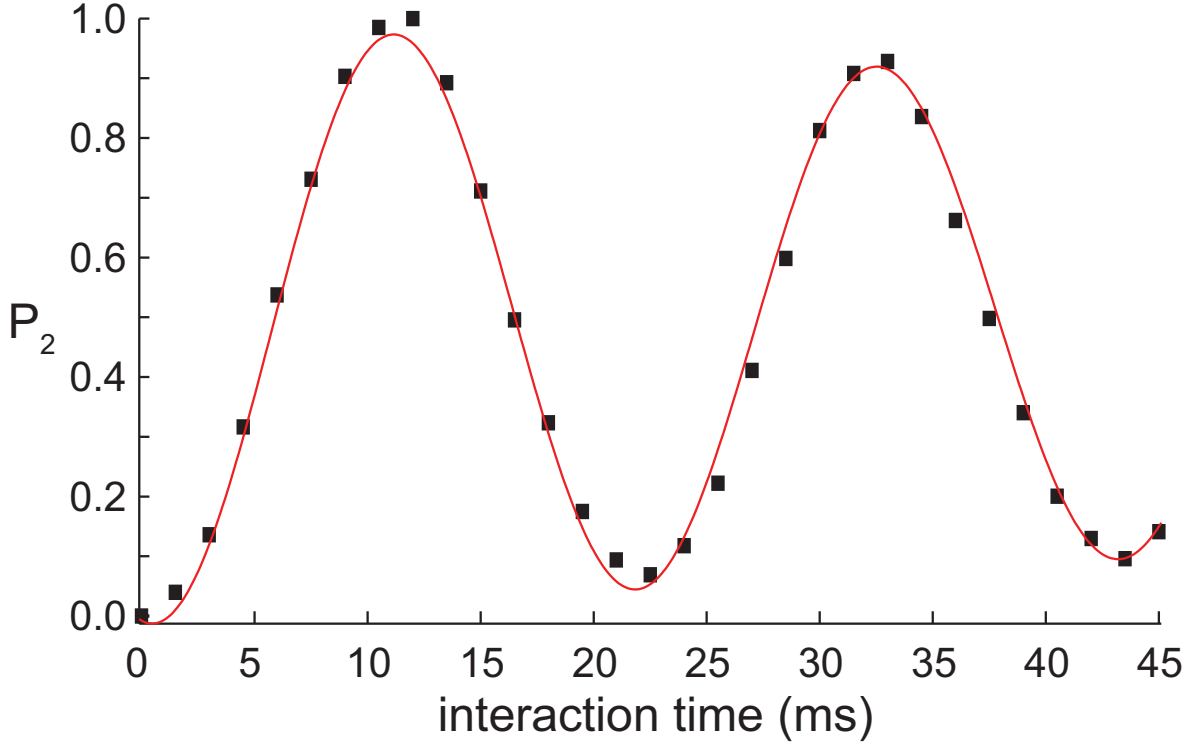}
\caption{\label{coheinf} Probability $P_2$  of the atoms undergoing M1 Rabi oscillations. The decay shows the long coherence time in the dipole trap with a blue detuning of 10.3 THz 
from the $D_2$ line. The red line is a fitting curve to the decaying Rabi oscillations, Eq.~\ref{eq:intanapole}.}
\end{figure}

Figure~\ref{coheinf} shows a Rabi oscillation measurement and its coherence time by looking at the number of atoms left in the upper state of the hyperfine manifold in the clock transition. We fit the data using Eq.~\ref{eq:intanapole} and extract a Rabi frequency of $2\pi\times$46.8~rad/s, as well as a decay time of 180 (30)~ms. We observe a linear relation between the decoherence rate and the Rabi frequency, which could be explained by the imperfect control of the external magnetic field (fluctuation peak to peak about 10 mG) or fluctuations of the microwave signal phase during its propagation from the source to the atoms due to imperfect connections.

\subsection{Sensitivity test}
Equation~\ref{signal} shows that $S$ has the largest sensitivity when $A_{PC}t_R/\hbar=(n+1/2)\pi$ (with $n$ an integer), which corresponds to the point with $P_2=0.5$  in Fig.~\ref{coheinf}. This interference scheme not only amplifies the parity-violation signal, but it also reduces the requirements on the stability of the transition frequency. 

\begin{figure}
\centering
\includegraphics[width=0.95\linewidth]{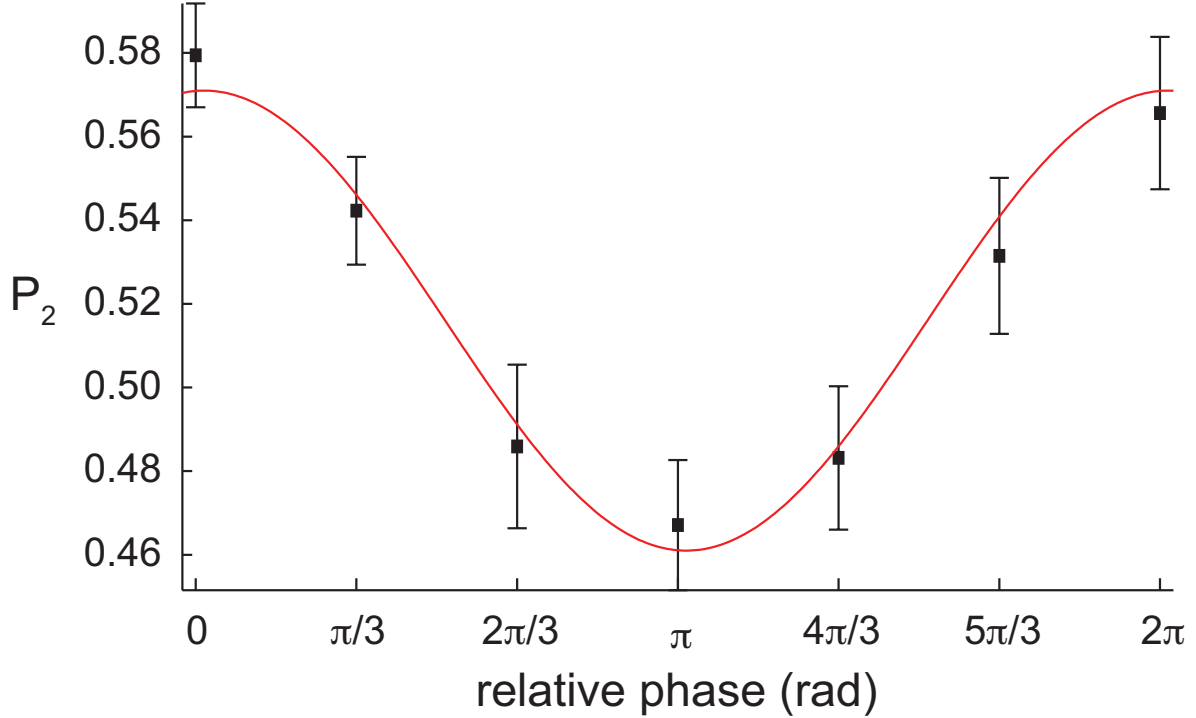}
\caption{\label{sensitivity}  Variation of $P_2$ at $t=37.5$ ms when the signal interferes with a second field having 100 smaller amplitude at the position of the atoms as a function of the phase between the two sources.}
\end{figure}

We test the sensitivity of the apparatus to a small change in the amplitude of the driving field as a way to measure the possibility of detecting the PNC signal. We use a second microwave source and horn to have independent control of the phase, which mimics the coordinate change, instead of simply attenuating the original drive by a minimum amount. In general, this additional interaction could be  any interaction that is phase-locked to the $M1$ transition driven by the first microwave horn, such as a stimulated Raman process. Our choice is based on the ability to control and calibrate our specific experimental apparatus and its parameters.
We proceed as follows:  start by performing the same experiment above using the second microwave source alone, and adjust the power to have the same Rabi frequency as that generated by the first microwave source alone (see Fig.~\ref{coheinf}). Then we connect  calibrated 40 dB power attenuators to the second source. This calibrates the microwave amplitude from the second source at the position of the atoms with respect to the first one.  Next we measure the effects using the interference method as we change the phase of the second microwave source. We choose the interaction time as $t_R=37.5$~ms, corresponding to $A_{PC}t_R/\hbar=7\pi/2$ and monitor the change in the excitation probability $P_2$ when tuning the phase of the second source. Fig.~\ref{sensitivity} shows the experimental results with the fitting curve, where the experimental data gives $P_2(\phi=0)-P_2(\phi=\pi)=0.11(1)$, while the prediction using the fitting parameters from Fig.~\ref{coheinf} gives $P_2(\phi=0)-P_2(\phi=\pi)=0.10(1)$. The fit gives an error to the amplitude of the oscillation of 0.004 so that our signal to noise ratio of this result is about 20 assuming that the noise is just the statistical uncertainty specified. We are not in the spin projection noise regime yet, on one hand due to the loss of atoms where the trap potential is not large compared with the atomic kinetic energy, and on the other hand because of the low collection efficiency of our imaging system ($<$2\%).

The amplitude of the second interfering microwave source is equivalent to a Rabi frequency of about 0.5~Hz, which is still two orders of magnitude larger than the expected anapole signal from Eq.~\ref{e1}. Future improvements include one more order-of-magnitude increase of the coherence time with further detuning and larger power of the trap laser, as well as a better light collection efficiency of the imaging system. We should stress once more that there has to be exquisite control of the apparatus and its environment to reach the required sensitivity (see Refs.~\cite{gomez07,sheng10}).

\section{Conclusion}
We have studied the differential ac Stark shift and coherence properties of a blue-detuned dipole trap for precision measurements by measuring Rabi oscillations of ground state superpositions in $^{87}$Rb. When the detuning is 20 nm from the $D_2$ line  we reach 180 ms of coherence lifetime with a Rabi frequency of the driving field of $2\pi\times 46.8$ rad/s and observe a differential ac Stark shift of 18 Hz. We have successfully implemented an interference between two phased microwave sources using the atoms as the detector with very different amplitudes. This method calibrates the sensitivity of our apparatus  to PNC-like signals. The result is that the interference amplifies the small signal, making it visible as a function of its phase. The trap is robust and shows strong promise for precision measurements, and the interference technique, with a tunable artificial amplitude, can be applied in other contexts to evaluate $S/N$ for precision experiments.

We thank Z. Kim and J. V. Porto for helpful discussions, and E. Gomez and J. A. Groover  for comments on the manuscript. This work is supported by NSF and DOE.

\end{document}